\documentstyle[twoside]{article}
\newcommand{\GeV}{\ifmmode\,{\rm GeV}\else~GeV\fi}
\newcommand{\MS}{\overline{\hbox{\sc ms}}}
\def\SU{{\rm SU}}

\makeatletter
\newcounter{alphaequation}[equation]
\def\thealphaequation{\theequation\alph{alphaequation}}
\def\eqnsystem#1{
\def\@eqnnum{{\rm (\thealphaequation)}}
\def\@@eqncr{\let\@tempa\relax
\ifcase\@eqcnt \def\@tempa{& & &}
\or \def\@tempa{& &}\or \def\@tempa{&}\fi\@tempa
\if@eqnsw\@eqnnum\refstepcounter{alphaequation}\fi
\global\@eqnswtrue\global\@eqcnt=0\cr}
\refstepcounter{equation}
\let\@currentlabel\theequation
\def\@tempb{#1}
\ifx\@tempb\empty\else\label{#1}\fi
\refstepcounter{alphaequation}
\let\@currentlabel\thealphaequation
\global\@eqnswtrue\global\@eqcnt=0
\tabskip\@centering\let\\=\@eqncr
$$\halign to \displaywidth\bgroup
  \@eqnsel\hskip\@centering
  $\displaystyle\tabskip\z@{##}$&\global\@eqcnt\@ne
  \hskip2\arraycolsep\hfil${##}$\hfil&
  \global\@eqcnt\tw@\hskip2\arraycolsep
  $\displaystyle\tabskip\z@{##}$\hfil
  \tabskip\@centering&\llap{##}\tabskip\z@\cr}

\def\endeqnsystem{\@@eqncr\egroup$$\global\@ignoretrue}
\makeatother

\begin{document}\large
\hfill\vbox{\baselineskip12pt
            \hbox{\bf IFUP -- TH 47/94}
            \hbox{\bf hep-ph/9411255}
            \hbox{October 1994}}

\begin{center}
\vglue 0.6cm{{\Large\bf\vglue 10pt
Light thresholds in\\ \vglue 3pt
Grand Unified Theories \\}
\vglue 1.0cm
{\large\bf Riccardo Barbieri, Paolo Ciafaloni and Alessandro Strumia\\[10mm] }
\baselineskip=13pt
{\em Dipartimento di Fisica, Universit\`a di Pisa \\[2mm] }
\baselineskip=12pt
{\rm and\\[2mm]}
\baselineskip=12pt
{\em INFN, Sezione di Pisa, I-56126 Pisa, Italy\\ }}

\vglue 2cm
{\large\bf Abstract}
\end{center}

\vglue 0.3cm
{\rightskip=3pc
 \leftskip=3pc
\baselineskip=12pt
 \noindent\large
In a generic Grand Unified Theory with a relatively small dispersion
of the spectrum around the $Z$-boson and the unification masses, a connection
is established, exact at one loop level, between $M_Z$, $G_{\rm F}$,
$\alpha(M_Z)$ and the strong coupling constant $\alpha_3(M_Z)$.
At this level of precision, this avoids the logical and phenomenological
inconsistency of predicting $\alpha_3(M_Z)$ by means of the electroweak
couplings as extracted from the data in the Standard Model rather than
in the complete theory.
Attention is paid to the independence of the physical results from
regularization and/or renormalization schemes.\\
\indent As a particularly relevant example, the analysis is specialized to
the case of the Minimal Supersymmetric Standard Model, with emphasis on
light charginos and neutralinos.}

\thispagestyle{empty}\newpage\setcounter{page}{1}

\normalsize

\section{Introduction}
The observed unification of the strong and electroweak coupling constants in a
supersymmetric Grand Unified Theory is among the very few direct experimental
results (the only one at present?) with a clear interpretation in terms of a
non-standard theory of the strong and the electroweak
interactions~\cite{Dino}. Needless to
say, such an interpretation requires a number of strong theoretical hypotheses
about the theory in its super-high energy regime. These hypotheses, however,
can
be formulated in a clear way. We would summarize them as follows:
\begin{itemize}
\item[i)]
the dispersion of the spectrum of the theory around the unification scale
$M_{\rm G}$
does not distort the evolution of the couplings in a significant way;
\item[ii)]
equally negligible are the effects related to the proximity of
$M_{\rm G}$ to the Planck
scale $M_{\rm Pl}$, as possibly due to the presence of higher dimensional
operators, scaled by inverse powers of $M_{\rm Pl}$.
\end{itemize}
Once these hypotheses are made, the possibility to determine the effects of the
unification of the couplings in the low energy theory is only limited by the
level of knowledge of the low energy theory itself and can in principle be
pushed to an arbitrary level of accuracy.

Along these lines, a consistent treatment of all possible effects up to
two loops in the $\beta$-function for the running of the couplings and
including
one loop ``threshold corrections'' is given in the relevant
literature~\cite{Langa}. Recently,
even one loop non-logarithmic corrections
have been considered~\cite{Fara,Lynn}, that scale like
powers of $M_Z/m_{\rm S}$, the ratio between the $Z$-mass and the mass of
any particle present in the low-energy spectrum.
As long as one expects some supersymmetric particle
with a mass comparable to $M_Z$, this is both logically
necessary and, maybe, phenomenologically relevant. The numerical results of
ref.~\cite{Fara} indicate that this can be the case.

A common feature of all these works, however, is the assumption of an initial
condition for two of the three relevant couplings at $M_Z$, obtained from a fit
of the various experimental data {\em inside\/} the Standard Model.
More explicitly, to
predict the strong coupling constant $\alpha_3(M_Z)$, one takes a value
for the electroweak couplings $\alpha_1 (M_Z)$ and
$\alpha_2 (M_Z)$ or, alternatively, for $\alpha(M_Z)$ and
$\sin^2\theta_{\rm W}(M_Z)$
using a Standard Model fit of the electroweak precision data. Clearly, in so
far as $M_Z\approx m_{\rm S}$, one makes in this way an error of the same
order of the effects that are being included by a proper treatment of the
thresholds corrections in the evolution of the couplings. Here we would like to
remedy to this shortcoming of the present analysis, by replacing the inputs for
$\alpha_1(M_Z)$ and $\alpha_2(M_Z)$ with direct experimental quantities.

In principle, thinking of all the different electroweak precision observables,
this would seem to require a complicated fitting procedure involving all of
them. In practice, this is not the case. A unified theory, with a given
dispersion of the spectrum around $M_Z$ and $M_{\rm G}$,
can be viewed as dependent on
three parameters (suitably defined): the grand scale, the low energy scale and
the unified coupling. As such, three measurements (``basic observables'') are
required to fix the theory and a fourth one (at least) to test it. As is well
known, for reasons of experimental precision and theoretical cleanliness, the
three basic observables that emerge are: $M_Z$, the Fermi constant $G_{\rm F}$,
as measured in $\mu$ decay,
and the electromagnetic fine structure constant at the $Z$ mass $\alpha(M_Z)$.
On the other hand,
any observable with a significant dependence on the strong coupling
can serve as a test of the unified theory. Since none, in this case, clearly
dominates as yet over the others, we shall take $\alpha_3(M_Z)$ itself,
defined in the
$\MS$ scheme, as commonly done in the literature.
We will therefore
establish a connection between $M_Z$, $G_{\rm F}$, $\alpha(M_Z)$
and $\alpha_3(M_Z)$
as function of the particle spectrum, taking into account all one loop
effects in an exact way. Needless to say, the particle spectrum will have
to be consistent with all data so far,
including the electroweak precision tests. In the
case of the supersymmetric particles,
once the constraints from production experiments are
satisfied, the virtual effects are sufficiently small to be generally
consistent
with observations~\cite{Alta}.
This is certainly the case if the
stop-sbottom splitting is not too large relative
to the mean mass and a tiny region of the parameter space
where charginos have a mass of $45\div 50\,\GeV$ is perhaps also avoided.
Notice, on the other hand,
that this is not an a priori prejudice to the possible relevance of
the considerations developed in this work, due to the different quality of the
determinations of $M_Z$, $G_{\rm F}$ and $\alpha(M_Z)$ relative to all other
electroweak
precision observables.

\section{General reference formulae}
We aim at establishing in this section the general connection between  $M_Z$,
$G_{\rm F}$, $\alpha(M_Z)$ and $\alpha_3(M_Z)$. Following Weinberg~\cite{Wein},
the program for calculating $\alpha_3(M_Z)$ to the required accuracy proceeds
in three steps, as follows:
\begin{itemize}
\item[a)]
Determine the renormalized $\SU(3)\otimes\SU(2)\otimes{\rm U}(1)$ couplings
$\alpha_i(\mu_{\rm H})$, $i=1,2,3$ at a scale $\mu_{\rm H}$ (H~=~heavy)
such that
$\mu_{\rm H}^2/M_{\rm G}^2\ll 1$, but still
$\ln(\mu_{\rm H}/M_{\rm G})={\cal O}(1)$.
Denoting by $\delta\alpha_i(\mu_{\rm H})$
the one loop contributions from the heavy particles to
the vacuum polarizations of the light gauge fields, we write, in terms of the
unified bare coupling $\alpha_{\rm G}$
\begin{equation}\label{eq:1}
\frac{1}{\alpha_i(\mu_{\rm H})}=\frac{1}{\alpha_{\rm G}}-
\frac{\delta\alpha_i(\mu_{\rm H})}{\alpha_i^2}.
\end{equation}
\item[b)]
By means of the two loop $\beta$-function for the $\SU(3)\otimes\SU(2)
\otimes{\rm U}(1)$ gauge group, run the couplings down to a scale
$\mu_{\rm L}$ (L~=~light) such that
$m_{\rm S}^2/\mu_{\rm L}^2\ll 1$, but $\ln(m_{\rm S}/\mu_{\rm L})={\cal O}(1)$,
where $m_{\rm S}$ is a typical light particle mass.
In this running $m_{\rm S}$ can be neglected. One can
write
\begin{equation}\label{eq:2}
\frac{1}{\alpha_i(\mu_{\rm L})}=\frac{1}{\alpha_i(\mu_{\rm H})}+
b_i \ln\frac{\mu_{\rm H}}{\mu_{\rm L}}+\delta_i^{\rm HL}
\end{equation}
where $b_i$ are the one loop $\beta$-function coefficients and
$\delta_i^{\rm HL}$ are the two loop terms
with $\mu_{\rm H}$ and $\mu_{\rm L}$ replaced, in a consistent
approximation, with $M_{\rm G}$ and $M_Z$ respectively. One has
\begin{equation}\label{eq:3}
\delta_i^{\rm HL}=\frac{1}{4\pi}\left[\frac{b_{it}^{(2)}}{6}
\ln\big(1+10.5\,\lambda_t^2(M_{\rm G})\big)+
\sum_{j=1}^3\frac{b_{ij}^{(2)}}{b_j}
\ln\frac{\alpha_j(M_{\rm G})}{\alpha_j(M_Z)}\right],
\end{equation}
where $b_{ij}^{(2)}$ are the two loop $\beta$-function coefficients
and $\lambda_t(M_{\rm G})$ is the top Yukawa coupling at the unification scale.
\item[c)]
{}From the inputs $\alpha_i(\mu_{\rm L})$, calculate the (suitably defined)
$\alpha_i(M_Z)$, taking into account all one loop effects of the light
particles
in $\delta\alpha_i(M_Z)$:
\begin{equation}\label{eq:4}
\frac{1}{\alpha_i(M_Z)}=\frac{1}{\alpha_i(\mu_{\rm L})}
-\frac{\delta\alpha_i(M_Z)}{\alpha_i^2}.
\end{equation}
\end{itemize}
{}From equations~(\ref{eq:1}),~(\ref{eq:2}),~(\ref{eq:4}),
it is a simple matter to obtain, by
eliminating $\alpha_{\rm G}$ and $\ln(\mu_{\rm H}/\mu_{\rm L})$,
\begin{equation}\label{eq:5}
\frac{1}{\alpha_3(M_Z)}=\frac{1}{b_{12}}(\frac{b_{13}}{\alpha_2(M_Z)}-
\frac{b_{23}}{\alpha_1(M_Z)})+\delta_{\rm H}+\delta_{\rm HL}+\delta_{\rm L}
\end{equation}
where $b_{ij}\equiv b_i-b_j$ and
\begin{eqnsystem}{sys:dHLHL}
\delta_{\rm HL}\! &=& \frac{1}{b_{12}}
(b_{23}\delta^{\rm HL}_1 + b_{31}\delta^{\rm HL}_2 + b_{12}\delta^{\rm
HL}_3),\\
\delta_{\rm H} &=&
-\frac{1}{b_{12}}\left[ b_{23}\frac{\delta\alpha_1(\mu_{\rm H})}{\alpha_1^2}
+ b_{31}\frac{\delta\alpha_2(\mu_{\rm H})}{\alpha_3^2}+
 b_{12}\frac{\delta\alpha_3(\mu_{\rm H})}{\alpha_3^2}\right],\\
\delta_{\rm L} &=&
-\frac{1}{b_{12}}\left[ b_{23}\frac{\delta\alpha_1(M_Z)}{\alpha_1^2}
+ b_{31}\frac{\delta\alpha_2(M_Z)}{\alpha_3^2}+
 b_{12}\frac{\delta\alpha_3(M_Z)}{\alpha_3^2}\right].
\end{eqnsystem}
Equation~(\ref{eq:5}) is not yet what we want, however.
To express $\alpha_3(M_Z)$
in terms of $M_Z$, $G_{\rm F}$ and $\alpha(M_Z)$, we need to connect them to
$\alpha_1(M_Z)$ and $\alpha_2(M_Z)$. Defining, as usual in the literature, an
auxiliary mixing angle parameter through
\begin{equation}
s^2\equiv \frac{1}{2}
\left[ 1-\sqrt{1-\frac{4\pi\alpha(M_Z)}{\sqrt{2} G_{\rm F}
M_Z^2}}\right],\qquad
c^2\equiv1-s^2,
\end{equation}
such connection, exact at one loop, is given by
\begin{eqnsystem}{sys:ai(GMa)}
\alpha_1(M_Z) &=& \frac{\alpha}{c^2}
\left[1-
\frac{\delta \alpha}{\alpha}+\frac{\delta c^2}{c^2}+
\frac{\delta\alpha_1(M_Z)}{\alpha_1}\right]\frac{5}{3}\\
\alpha_2(M_Z) &=&
\frac{\alpha}{s^2}\left[1-
\frac{\delta \alpha}{\alpha}+\frac{\delta s^2}{s^2}+
\frac{\delta\alpha_2(M_Z)}{\alpha_2}\right],
\end{eqnsystem}
where
$$\delta s^2=-\delta c^2=\frac{s^2c^2}{c^2-s^2}
\left[\frac{\delta \alpha}{\alpha}-\frac{\delta G_{\rm F}}{G_{\rm F}}-
\frac{\delta M_Z^2}{M_Z^2}\right]$$
and $\delta\alpha$, $\delta G_{\rm F}$, $\delta M_Z^2$
are the full one loop corrections from all the light
particles to $\alpha(M_Z)$, $G_{\rm F}$ and $M_Z^2$
respectively. This
allows us to write down the result for $\alpha_3(M_Z)$ in the desired form
\begin{equation}\label{eq:Predizione}
\frac{1}{\alpha_3(M_Z)}=
\frac{b_{13} s^2 - \frac{3}{5} b_{23}c^2}{b_{12}\alpha(M_Z)}
+\delta_{\rm H}+\delta_{\rm HL}+\Delta_{\rm L},
\end{equation}
where, in this case, the light particle contribution to the one loop correction
$\delta_{\rm L}$ is replaced by
\begin{equation}\label{eq:11}
\Delta_{\rm L} = -
\frac{1}{b_{12}}\left[ b_{23}\frac{\Delta\alpha_1(M_Z)}{\alpha_1^2}
+ b_{31}\frac{\Delta\alpha_2(M_Z)}{\alpha_3^2}+
 b_{12}\frac{\delta\alpha_3(M_Z)}{\alpha_3^2}\right]
\end{equation}
\begin{eqnsystem}{sys:Deltai(GMa)}
\frac{\Delta \alpha_1(M_Z)}{\alpha_1} &=&
\frac{\delta\alpha}{\alpha}-\frac{\delta c^2}{c^2},\\
\frac{\Delta \alpha_2(M_Z)}{\alpha_2} &=&
\frac{\delta\alpha}{\alpha}-\frac{\delta s^2}{s^2}.
\end{eqnsystem}
Finally, $\Delta\alpha_1(M_Z)$ and $\Delta\alpha_2(M_Z)$ are easily expressed
in terms of commonly defined amplitudes~\cite{Barbieri}. One has
\begin{eqnsystem}{sys:Deltai(pol)}
\frac{\Delta\alpha_1(M_Z)}{\alpha_1} &=&-
F_{00}(M_Z^2) +\frac{1}{c^2-s^2}
\left[s^2 (e_1-\frac{\delta G_{\rm VB}}{G_{\rm F}})
-(2s^2 e_3+2\frac{s}{c}\frac{A_{\gamma Z}}{M_Z^2})-c^2 e_4\right]
\label{eq:13a}\\
\frac{\Delta\alpha_2(M_Z)}{\alpha_2} &=&-
F_{33}(M_Z^2) -\frac{1}{c^2-s^2}
\left[c^2 (e_1-\frac{\delta G_{\rm VB}}{G_{\rm F}})
-(2s^2 e_3+2\frac{s}{c}\frac{A_{\gamma Z}}{M_Z^2})-s^2 e_4\right]\label{eq:13b}
\end{eqnsystem}
where, from the vacuum polarization amplitudes
($i,j=W,Z,\gamma$ and $i,j=0,3$ for $B,W_3$)
$$\Pi_{\mu\nu}^{ij}(q^2) = -i g_{\mu\nu}[A_{ij} + q^2 F_{ij}(q^2)]+
q_\mu q_\nu~{\rm terms},$$
\begin{eqnarray*}
e_1&\equiv&\frac{A_{33}-A_{WW}}{M_W^2}=\frac{A_{ZZ}}{M_Z^2}-
\frac{A_{WW}}{M_W^2}+2\frac{s}{c}\frac{A_{\gamma Z}}{M_Z^2}\\
e_3&\equiv&\frac{c}{s} F_{03}(M_Z^2)\\
e_4&\equiv& F_{\gamma\gamma}(0)-F_{\gamma\gamma}(M_Z^2)
\end{eqnarray*}
and $\delta G_{\rm VB}$ is the one loop correction, except for vacuum
polarizations (vertices, boxes, and fermion self-energies), to
the $\mu$-decay amplitude~\cite{Barbieri}.
Notice that $e_4$ contains all the one loop contributions to the photon
vacuum polarization which are not included in $\alpha(M_Z)$.
Conventionally, we take $\alpha(M_Z)$ to contain only the corrections
from all leptons and quarks except the top.

Before going to numerical results, we find it useful to make clear
to what extent the final result, eq.~(\ref{eq:Predizione}),
is independent from regularization and/or renormalization schemes.
If, in place of $\alpha_3(M_Z)$, we had used a direct
physical observable, like, e.~g., the hadronic width of the
$Z$-boson, no reference to any regularization or renormalization scheme
should have been made at all, except, possibly,
in the intermediate steps of the calculation.

The first term in the right-hand side of eq.~(\ref{eq:Predizione}), dependent
on physical observables only, has an absolute meaning. The heavy particle
correction term $\delta_{\rm H}$ has also an intrinsic definition, being
obtained through the gauge invariant~\cite{Wein} one loop contributions to
the vacuum polarizations of the light gauge bosons.
Since $(\mu_{\rm H}/M_{\rm G})^2\ll 1$,
$\delta_{\rm H}$ does not depend on $\mu_{\rm H}$ and, because of unification,
it is ultraviolet finite. In spite of that, it does however depend upon the
regularization scheme. In terms of the masses $\{M_h\}$ of the heavy
particles and their contributions $b_i^h$ to the $\beta$-functions
coefficients $b_i$ for the three gauge group factors
in $\SU(3)\otimes\SU(2)\otimes{\rm U}(1)$, one has
\begin{equation}\label{eq:dh}
\delta_{\rm H}=\frac{1}{2\pi b_{12}}\sum_h(b_{23}b_1^h+
b_{31}b_2^h+b_{12}b_3^h)\ln\frac{M}{M_h}+
\frac{2b_{13}-3b_{12}}{12\pi b_{12}},
\end{equation}
with the mass independent term given in the usual $\MS$ scheme. Note also that
$\delta_{\rm H}$ does not depend on the overall scale $M$ since, due to
unification,
\begin{equation}
\sum_h b_i^h=b_{\rm G}-b_i.
\end{equation}
where $b_{\rm G}$ is the one loop coefficient of the $\beta$-function
for the unified group.

Unlike $\delta_{\rm H}$, the light particle correction
term $\Delta_{\rm L}$ does not
have an intrinsic definition, but depends on the definition of $\alpha_3(M_Z)$
via $\delta\alpha_3(M_Z)$. We stick to the usual $\MS$ definition of
$\alpha_3(M_Z)$, so that
\begin{equation}\label{eq:delta3}
\frac{\delta\alpha_3(M_Z)}{\alpha_3^2}=-\frac{b_3}{4\pi}
(\frac{2}{d-4}-\gamma_{\rm E}+\ln4\pi+\ln \frac{M_Z^2}{\mu_{\rm MS}^2})
+\sum_\ell\frac{b_3^\ell}{4\pi}\ln \frac{M_\ell^2}{M_Z^2},
\end{equation}
where $d$ is the dimension of space time,
$\mu_{\rm MS}$ is the dimensional regularization mass scale, and the sum
extends over all the light particles heavier than $M_Z$.
The calculation of $\Delta_{\rm L}$ is completed by the expressions
for $\Delta\alpha_1(M_Z)$ and $\Delta\alpha_2(M_Z)$, which are
intrinsically defined in any given theory.
The overall $\Delta_{\rm L}$, although ultraviolet finite, has however
a regularization dependence which would compensate the regularization
dependence of $\delta_{\rm H}$, if we had computed the
strong interaction corrections
to a physical observable instead of $\alpha_3(M_Z)$ itself.
The compensation comes about as follows.
For example in dimensional regularization, different extensions of
the Lorentz and Dirac indices
(algebra of $\gamma$ matrices in $d$ dimensions,
definition of $\gamma_5$, etc.)
amounts to obtain a finite contribution to the
various $\delta\alpha_i$ of the form
\begin{equation}\label{eq:deltaReg}
\frac{\delta_{\rm reg}\alpha_i}{\alpha_i^2} =
b_{i\rm V} f_{\rm V} + b_{i\rm F} f_{\rm F} + b_{i\rm S} f_{\rm S}
\end{equation}
where $b_{i\rm V},b_{i\rm F},b_{i\rm S}$ are the contributions
to the $\beta$-function
coefficients of the Vector, the Fermions and the Scalars respectively,
whereas $f_{\rm V},f_{\rm F}$ and $f_{\rm S}$ are constants dependent
on the specific regularization scheme~\cite{Langa,Wein,Anto}.
If eq.~(\ref{eq:deltaReg}) is inserted in all $\delta\alpha_i$, or
$\Delta\alpha_i$, both in $\delta_{\rm H}$ and $\Delta_{\rm L}$, a full
cancellation takes place for any $f_{\rm V},f_{\rm F},f_{\rm S}$
since the vectors,
the fermions and the scalars, light and heavy, form individual
complete multiplets of the unified group.
The cancellation is incomplete, because $\delta\alpha_3(M_Z)$ is defined
in~(\ref{eq:delta3}) without the extra regularization dependent term
in~(\ref{eq:deltaReg}), and is only recovered when the relation
between $\alpha_3(M_Z)$ and the physical observable is established.

\section{The case of the Minimal Supersymmetric Standard Model}
We can now specialize the result of eq.~(\ref{eq:Predizione}) to the case of
the Minimal Supersymmetric Standard Model~\cite{Nilles}.
The dominant term in the right hand side of eq.~(\ref{eq:Predizione}), using
\begin{eqnsystem}{sys:misure}
M_Z &=& 91.188\pm0.0044\GeV\\
G_{\rm F} &=& 1.16637(2)\cdot 10^{-5}\GeV^{-2} \\
\alpha(M_Z) &=&  (128.87\pm0.12)^{-1}\\
(s^2 & = &0.2312\pm0.0003)
\end{eqnsystem}
and
$$b_1=\frac{33}{5},\qquad b_2=1,\qquad b_3=-3,$$
gives
\begin{equation}
\alpha_3(M_Z){(\rm leading~logs)}=0.1155\pm0.0010.
\end{equation}
For the two loop term $\delta_{\rm HL}$, also including a correction involving
the top Yukawa coupling, one has~\cite{Langa}
\begin{equation}
\delta_{\rm HL}=-0.80\pm 0.10.
\end{equation}
Assuming no dispersion of the heavy particle spectrum, the heavy particle
contribution~(\ref{eq:dh}) gives a negligible effect
($\delta_{\rm H}\approx 0.01$), so that, before the inclusion of
$\Delta_{\rm L}$, $\alpha_3 = 0.1273\pm0.0020$.

\begin{table}
$$\begin{array}{cc|cc}
m_t~{\rm in~GeV} & \Delta_{\rm L}^t(m_t) & m_h~{\rm in~GeV} &
\Delta_{\rm L}^h(m_h) \\ \hline
120 & -0.898 & ~50 & 1.117 \\
130 & -0.977 & ~60 & 1.140 \\
140 & -1.061 & ~70 & 1.160 \\
150 & -1.151 & ~80 & 1.177 \\
160 & -1.246 & ~90 & 1.193 \\
170 & -1.347 & 100 & 1.207 \\
180 & -1.452 & 110 & 1.220 \\
190 & -1.564 & 120 & 1.232 \\
200 & -1.680 & 130 & 1.243 \\
210 & -1.803 & 140 & 1.254 \\
220 & -1.903 & 150 & 1.263 \\
\end{array}$$
\caption{numerical values of
$\Delta_{\rm L}^t(m_t)$ and $\Delta_{\rm L}^h(m_h)$
for representative values of
$m_t$ and $m_h$.}
\end{table}

\begin{figure}[t]\setlength{\unitlength}{1cm}
\begin{center}\begin{picture}(16,6.5)
\put(0,0){\includegraphics{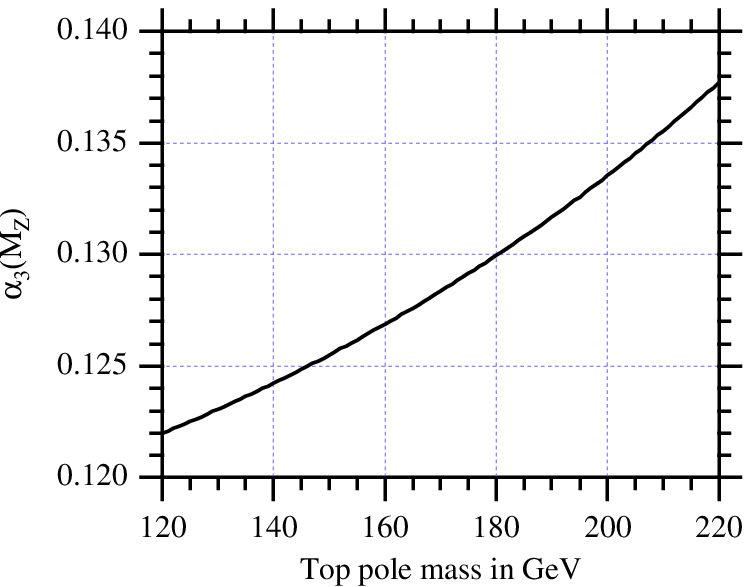}}
\put(8,0){\includegraphics{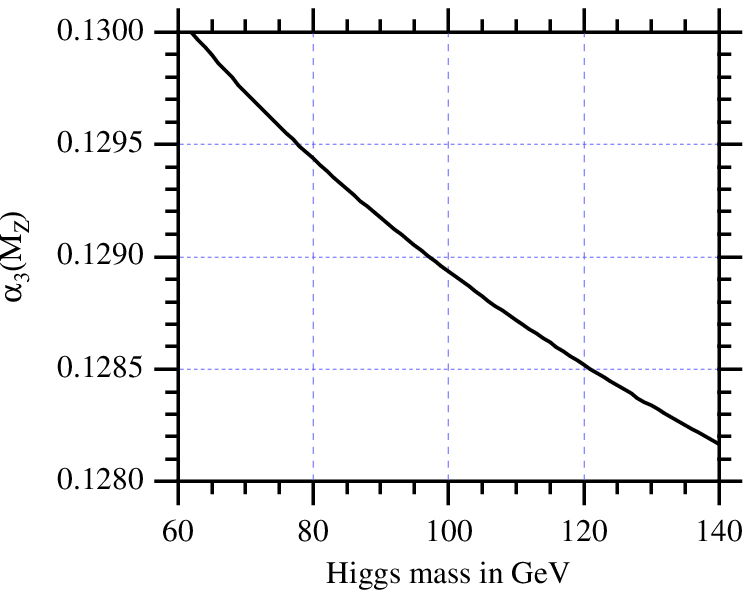}}
\put(4,6.2){Fig.~1a}
\put(12,6.2){Fig.~1b}
\end{picture}
\caption{$\alpha_3(M_Z)$ as function
of the top pole mass $M_t$ for $m_h=M_Z$ (1a) and
of $m_h$ for $M_t=175~{\rm GeV}$ (1b),
taking $\Delta_{\rm L}^{\rm susy}=0$ and $\delta_{\rm HL}=-0.80$.
\label{fig:a3(ht)}}
\end{center}\end{figure}

Coming to the light particle correction term $\Delta_{\rm L}$,
let us first consider
the case where all the extra particles introduced by supersymmetry are heavy
with respect to $M_Z$, so that all power law corrections in $M_Z/m_{\rm S}$
can be safely neglected. In this situation, apart from a contribution to the
cancellation of the divergences, the supersymmetric particles only contribute
to $\Delta_{\rm L}$ with logarithmic terms, $\ln (m_{\rm S}/M_Z)$,
that have been studied in the literature~\cite{Langa,Ross}.
On the other hand, also the various contributions to
$e_1$, $e_3$, $e_4$, $F_{00}(M_Z)$, $F_{33}(M_Z)$ and $\delta G_{\rm VB}/G_{\rm
F}$
from $W$, $Z$, $\gamma$, light fermions, top and standard
Higgs boson, can be reconstructed from the literature~\cite{Fij}.
By putting everything together, taking all supersymmetric particles
degenerate at $m_{\rm S}$, we find, for $m_t>M_Z/2$,
\begin{eqnsystem}{sys:DLhtSUSY}
\Delta_{\rm L} &=& \Delta_{\rm L}^t(m_t)+\Delta_{\rm L}^h(m_h)+
\Delta_{\rm L}^{\rm susy}\\
\Delta_{\rm L}^t  &=&\frac{45}{112\pi}\frac{1}{c^2-s^2} \Bigg\{
{4\over 81}(16s^4+128s^2-73)
-t - \frac{2}{9}\ln t -\\
 &&+ \frac{4}{27}[(24s^2-9-32s^4)+(9+48s^2-64s^4)t]
\left[1-\sqrt{4t-1}\arcsin\frac{1}{\sqrt{4t}}\right] \Bigg\}\nonumber\\
\Delta_{\rm L}^h  &=& 1.197 +\frac{45}{112\pi}\frac{1}{c^2-s^2} \Bigg\{
\frac{h^2 s^2\ln h}{(1-h)(h-c^2)}+
\frac{hc^2\ln c^2 }{h-c^2}+\frac{h}{18}+  \\
&& +\frac{1}{9}(12-4h+h^2)\left[1 + (\frac{h}{h-1}-\frac{h}{2})
\ln h-h\sqrt{{4\over h}-1}\arctan\sqrt{{4\over h}-1}\right]
 \Bigg\} \nonumber \\
\Delta_{\rm L}^{\rm susy} &=&
\frac{19}{28\pi}\ln\frac{m_{\rm S} }{M_Z}\qquad
\hbox{for $m_{\rm S}^2\gg M_Z^2$},
\end{eqnsystem}
where $t\equiv m_t^2/M_Z^2$ and $h\equiv m_h^2/M_Z^2$.
The numerical values of $\Delta_{\rm L}^t$ and $\Delta_{\rm L}^h$
for representative values of
$m_t$ and $m_h$ are given in table~1.
As overall result, the prediction for $\alpha_3(M_Z)$
as function of $m_t$ and
$m_h$, taking $\delta_{\rm H} = 0$, $\delta_{\rm HL}=-0.80$ and,
for sake of illustration, $\Delta_{\rm L}^{\rm susy}=0$
is shown in fig.~\ref{fig:a3(ht)}.

Of course our interest is in possibly significant deviations from these
predictions for $\alpha_3(M_Z)$ coming from light supersymmetric particles
like,
e.g., gaugino-higgsinos or sfermions. On the contrary, for the purpose of the
present discussion, we can safely neglect possible deviations in the Higgs
system from the case of the Standard Model with a light Higgs. Let us
consider a particular
set of light supersymmetric particles. The corresponding contribution to
$\Delta_{\rm L}$, exact to one loop, can be obtained
from eq.s~(\ref{eq:11},\ref{eq:13a},\ref{eq:13b},\ref{eq:delta3}). We find
\begin{eqnarray}
\label{eq:25}
\Delta_{\rm L}^{\rm susy}(\hbox{light sparticles}) &=&
42.46(F_{00}^\ell-\frac{\alpha_1}{4\pi}b_1^\ell \Delta_Z)-
51.07(F_{33}^\ell-\frac{\alpha_2}{4\pi}b_2^\ell \Delta_Z)+ \\
&&-91.29(e_1^\ell+\frac{\delta G_{\rm VB}^\ell}{G_{\rm F}})+
80.44 e_3^\ell+82.68 e_4^\ell-
\sum_{\ell}\frac{b_3^\ell}{4\pi}\ln\frac{M_\ell^2}{M_Z^2}\nonumber,
\label{eq:26}\end{eqnarray}
where the sum extends over all the light sparticles heavier than $M_Z$,
$$\Delta_Z\equiv \frac{2}{d-4}-\gamma_{\rm E}+\ln 4\pi+
\ln\frac{M_Z^2}{\mu^2_{\rm MS}},$$
$b_i^\ell$ is the contribution to the
$\beta$-function coefficients of the light supersymmetric particles
and we denote by $F_{00}^\ell$, $F_{33}^\ell$,
$\delta G_{\rm VB}^\ell$, $e_1^\ell$,
$e_3^\ell$, $e_4^\ell$ their contributions to the corresponding functions
and by $M_\ell$ their masses.
The divergent pieces in
$\Delta_{\rm L}^{\rm susy}$
have also been properly subtracted away.

\begin{figure}[tp]\setlength{\unitlength}{1cm}
\begin{center}\begin{picture}(16.5,21)
\end{picture}
\put(-17.5,9){\includegraphics{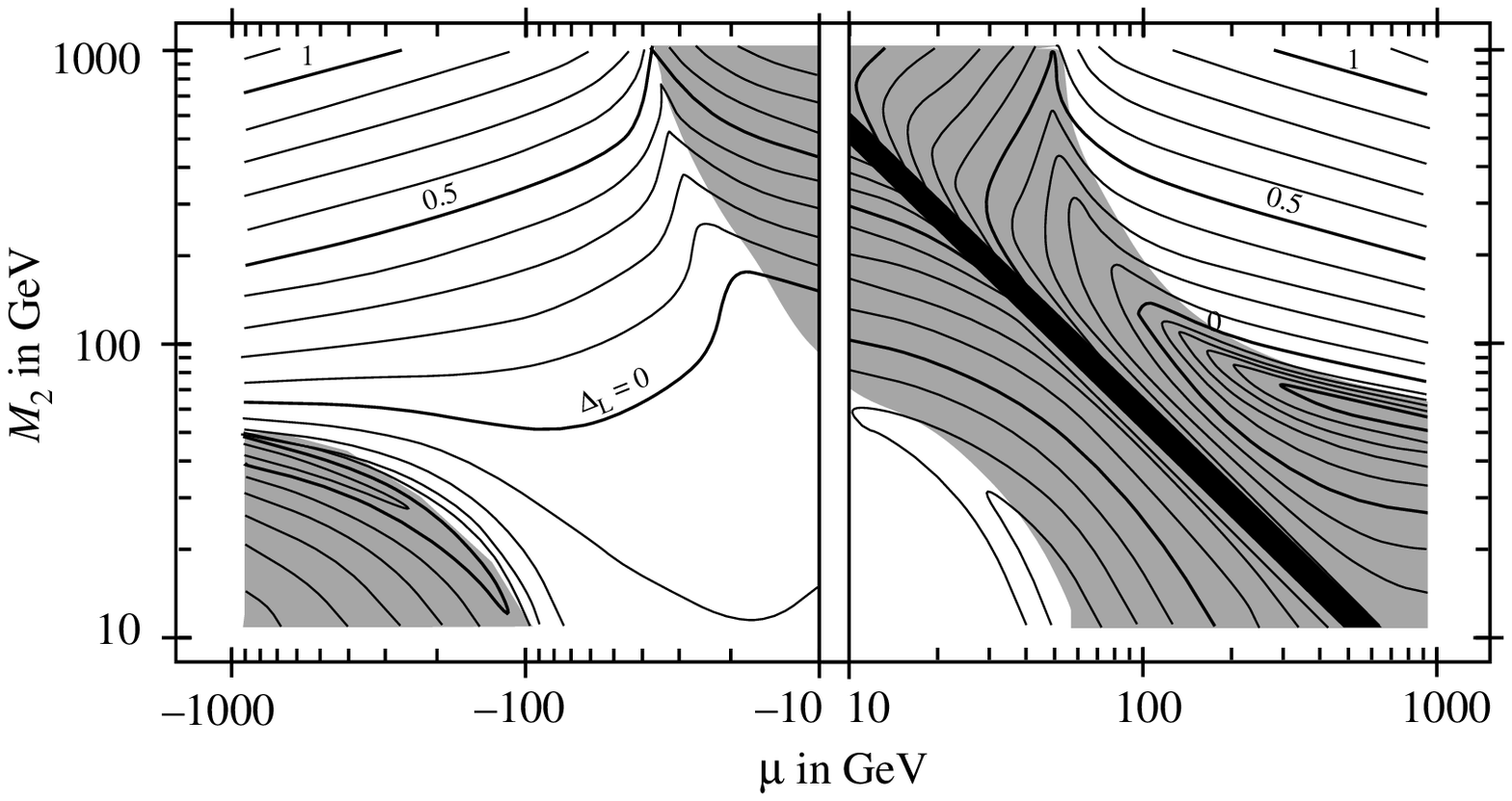}}
\put(-17.5,-2){\includegraphics{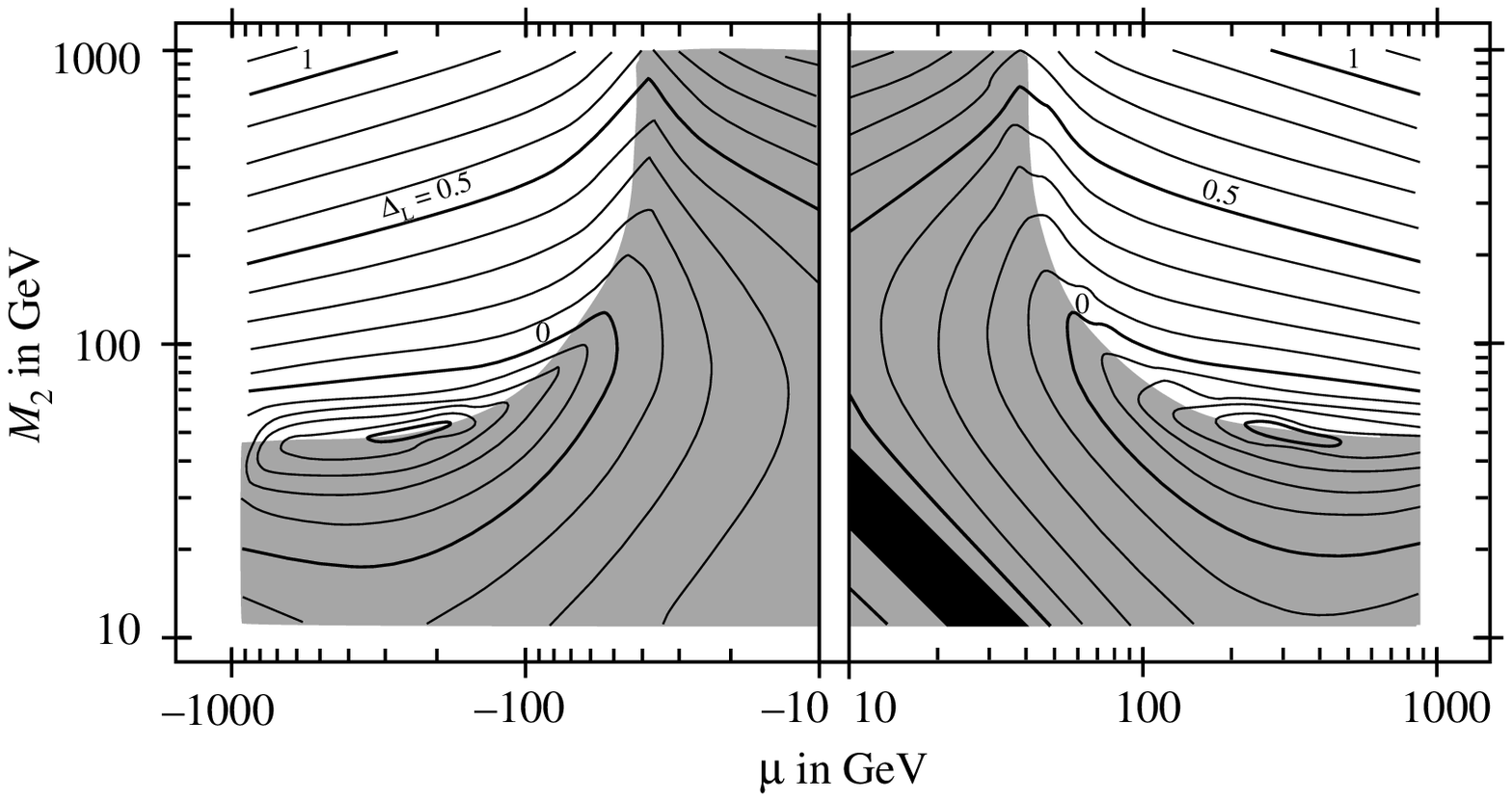}}
\put(-9,20){Fig.~2a: $\tan\beta=1.5$}
\put(-9,9){Fig.~2b: $\tan\beta=40$}
\caption{contribution to $\Delta_{\rm L}$
from charginos and neutralinos for $\tan\beta=1.5$ (fig.~2a)
and $\tan\beta=40$ (fig.~2b). The darker region corresponds to
a chargino lighter than 45~GeV.\label{fig:dL}}
\end{center}\end{figure}

\begin{figure}\setlength{\unitlength}{1cm}
\begin{center}\begin{picture}(16.5,21)
\end{picture}
\put(-17.5,9){\includegraphics{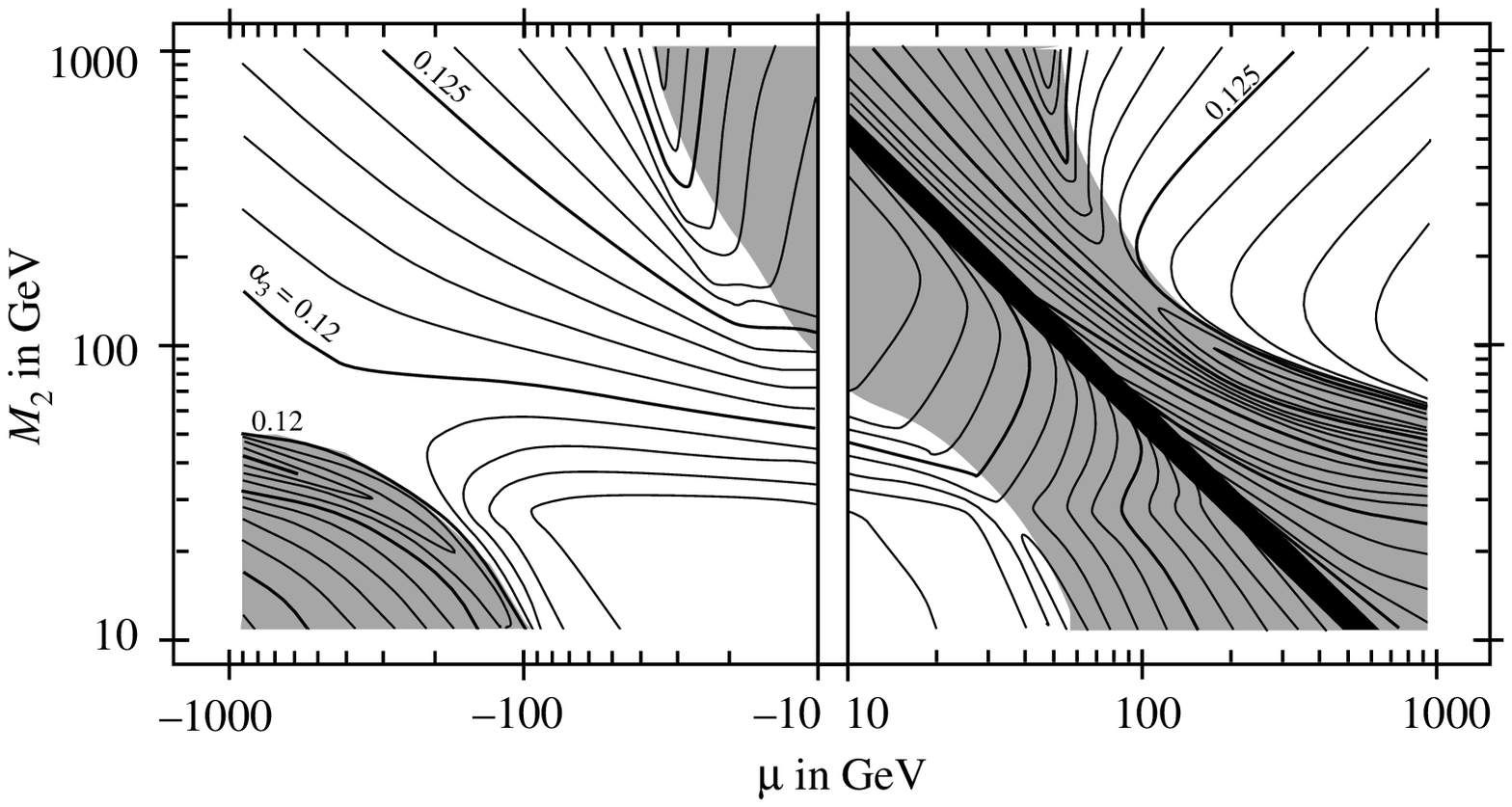}}
\put(-17.5,-2){\includegraphics{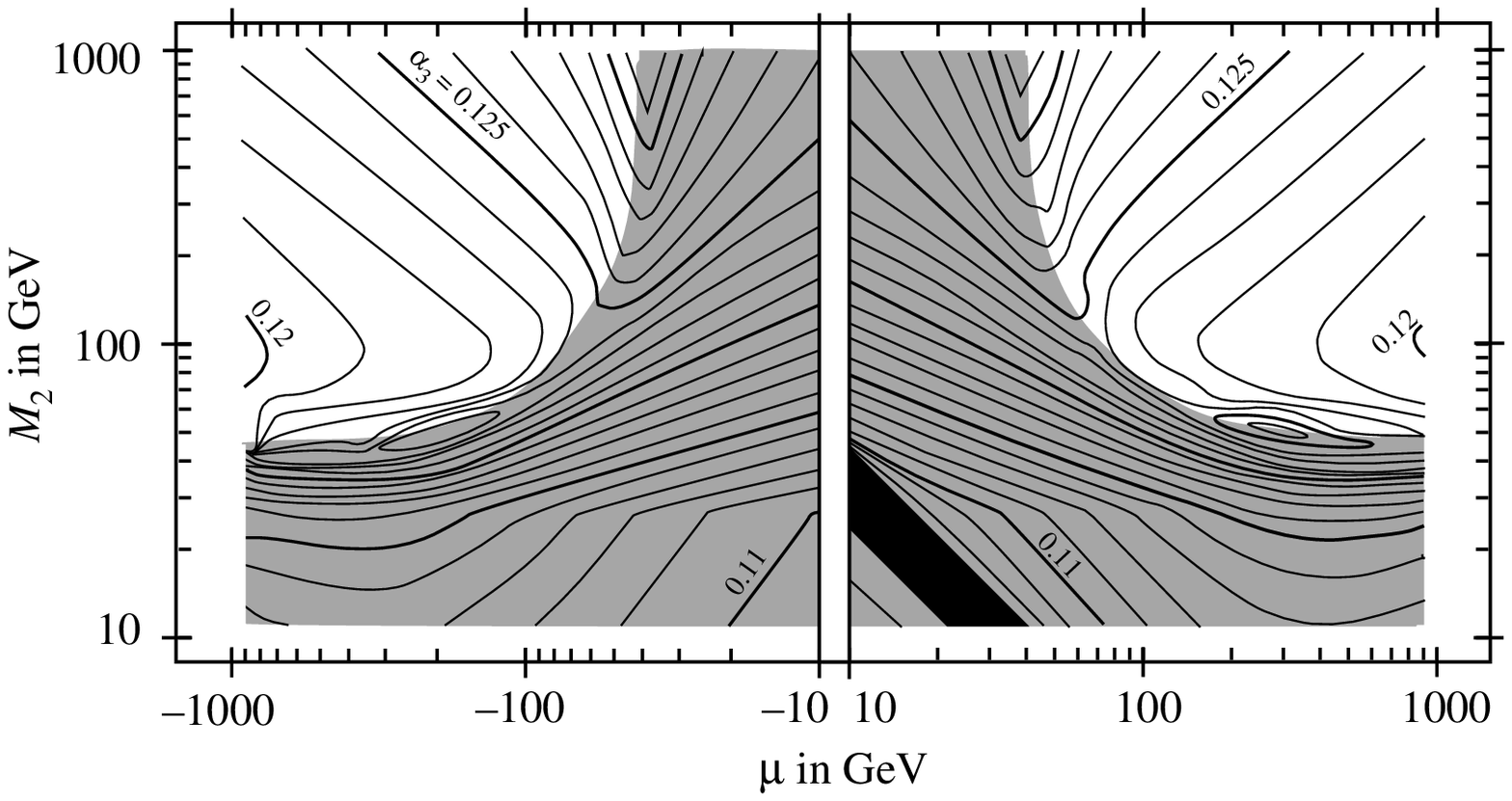}}
\put(-9,20){Fig.~3a: $\tan\beta=1.5$}
\put(-9,9){Fig.~3b: $\tan\beta=40$}
\caption{$\alpha_3(M_Z)$ in the plane $(\mu,M_2)$ for $\tan\beta=1.5$
(fig.~3a) and $\tan\beta=40$ (fig.~3b). Here the top pole mass is
$M_t=175$~GeV, $m_h=M_Z$,
$m({\rm gluino})=3.6\,M_2$ and, from sfermions and heavy Higgs,
$\Delta_{\rm L}=1$.
\label{fig:a3}}
\end{center}\end{figure}

The value of $\Delta_{\rm L}^{\rm susy}$ can be analyzed in
the parameter space of the Minimal Supersymmetric Standard Model.
We find that it is most likely to be relevant in the case of light
charginos and neutralinos, illustrated in figures~\ref{fig:dL},\ref{fig:a3}.
In figure~\ref{fig:dL} we give in the usual parameter space~\cite{Nilles}
the contribution
to $\Delta_{\rm L}$ from charginos and neutralinos,
for different values of $\tan\beta$.
In the logarithmic scale that we use, the deviation from the purely
logarithmic approximation distorts the contour plot from a pure
straight and equidistant line pattern.
Such distortion is significant for values of $\mu$ and $M_2$
(as renormalized at low energy) reasonably close to the
boundary of the excluded region, defined by requiring the lightest chargino
to be heavier than 45~GeV.
Numerically the deviation from the logarithmic approximation varies from
$-1.0$ to $0.15$ for $\tan\beta=1.5$ and from
$-0.65$ to $0.0$ for $\tan\beta=40$.
The relative size of this effect, from point to point in parameter space,
can be comparable to the two loop $\beta$-function correction $\delta_{\rm
HL}$.

Always with the focus on possibly light charginos and neutralinos,
we have collected in figures~\ref{fig:a3} all the various contribution
to $\alpha_3(M_Z)$.
There we take $m_t=175$~GeV, $m_h=M_Z$,
the gluino mass as consistently determined from $M_2$ and
the usual unification condition, and, for all other heavy particles,
sfermions and heavy Higgs bosons, $\Delta_{\rm L}=1$.
Within the assumption stated in the introduction, and for fixed values
of the parameters as explained, the uncertainty in the value of
$\alpha_3(M_Z)$ given in fig.s~\ref{fig:a3} is about
$\Delta\alpha_3(M_Z)\approx\pm\,2\cdot10^{-3}$.

\nonfrenchspacing
\end{document}